\documentclass[11pt]{article}
\usepackage[dvipdfmx]{graphicx}
\usepackage{color}
\usepackage{amsmath,amssymb,amsfonts,latexsym}
\usepackage{makecell}
\usepackage{float}

\begin{document}

\begin{center}
\textbf{\Large Backward Smoothing versus Fixed-Lag Smoothing\\[2mm] in Particle Filters}

\vspace{8mm}
{\large Genshiro Kitagawa}\\[4mm]
Tokyo University of Marine Science and Techonology\\[-1mm]
and\\[-1mm]
The Institute of Statistical Mathematics

\vspace{3mm}
{\today}
\end{center}

\begin{abstract}
Particle smoothing enables state estimation in nonlinear and
non-Gaussian state-space models, but its practical use is often limited
by high computational cost.
Backward smoothing methods such as the Forward Filter Backward Smoother
(FFBS) and its marginal form (FFBSm) can achieve high accuracy, yet
typically require quadratic computational complexity in the number of
particles.
This paper examines the accuracy--computational cost trade-offs of
particle smoothing methods through a trend-estimation example.
Fixed-lag smoothing, FFBS, and FFBSm are compared under Gaussian and
heavy-tailed (Cauchy-type) system noise, with particular attention to
$\mathcal{O}(m)$ approximations of FFBSm based on subsampling and local
neighborhood restrictions.
The results show that FFBS and FFBSm outperform fixed-lag smoothing at a
fixed particle number, while fixed-lag smoothing often achieves higher
accuracy under equal computational time.
Moreover, efficient FFBSm approximations are effective for Gaussian
transitions but become less advantageous for heavy-tailed dynamics.
\end{abstract}

\vspace{2mm}

\section{Introduction}

State-space models provide a flexible and powerful framework for
analyzing time-series data by explicitly separating the latent system
dynamics from the observation mechanism.
When the model is linear and Gaussian, optimal filtering and smoothing
can be performed efficiently by the Kalman filter and fixed-interval
smoother (\cite{AM 1979}).
However, many real-world phenomena exhibit nonlinear dynamics and
non-Gaussian disturbances, for which closed-form solutions are no
longer available (\cite{KG 1996}, \cite{Kitagawa 2020}).

Sequential Monte Carlo methods, commonly referred to as particle
filters and smoothers, were developed to address this limitation.
By representing probability distributions with a finite set of random
samples (particles), particle methods enable recursive filtering and
smoothing for general nonlinear and non-Gaussian state-space models.
Since their introduction, they have been successfully applied to a wide
range of problems in signal processing, econometrics, and engineering
(\cite{DFG 2001}, \cite{GSS 1993}, \cite{Kitagawa 1996}, \cite{Kitagawa 2020}).

Despite their flexibility, particle smoothers suffer from intrinsic
computational and statistical challenges.
In particular, smoothing algorithms that rely on backward recursions
often experience particle degeneracy, where only a small fraction of
particles effectively contribute to the approximation (\cite{DFG 2001}, \cite{Kitagawa 1996}).
Moreover, backward smoothing methods such as the Forward Filter
Backward Smoother (FFBS) and its marginal version (FFBSm) incur a high
computational cost, typically of order $\mathcal{O}(m^2)$ per time step,
which severely limits their practical applicability for large particle
numbers (\cite{BDM 2010}).

Several approaches have been proposed to alleviate these difficulties,
including fixed-lag smoothing, backward simulation, and various
$\mathcal{O}(m)$ approximations of FFBSm based on truncation, kernel
approximations, or subsampling.
Fixed-lag smoothing offers a computationally attractive alternative,
but introduces a bias that depends on the chosen lag length.
Approximate FFBSm methods aim to retain the statistical efficiency of
backward smoothing while reducing computational complexity, yet their
performance strongly depends on the structure of the state dynamics and
the tail behavior of the system noise (\cite{BDM 2010}, \cite{Kitagawa 1996}).

The purpose of this paper is to empirically investigate the
accuracy--computational cost trade-offs of particle smoothing methods
through a simple but informative trend-estimation example.
Focusing on a first-order trend model, we compare fixed-lag smoothing,
FFBS, and FFBSm under Gaussian and heavy-tailed (Cauchy-type) system
noise (\cite{KG 1996}, \cite{Kitagawa 1996}).
Special attention is paid to $\mathcal{O}(m)$ approximations of FFBSm
based on subsampling and local neighborhood restrictions, and to how
their effectiveness depends on the tail behavior of the transition
distribution.

Although the numerical study is limited to a single class of models,
it clearly illustrates the conditions under which fast particle
smoothing methods can be effective, and when simpler alternatives such
as fixed-lag smoothing are preferable under realistic computational
constraints.
The results provide practical guidance for selecting particle smoothing
strategies in applications where both accuracy and computational
efficiency are of concern.

The remainder of the paper is organized as follows.
Section~2 briefly reviews filtering and smoothing algorithms for
state-space models.
Section~3 describes particle filters and smoothers, including FFBS,
FFBSm, and their computationally efficient approximations.
Section~4 presents the numerical experiments and discusses the key
findings.
Section~5 concludes with a summary and practical implications.

\section{A Brief Review of the Filtering and Smoothing Algorithms}

\subsection{The state-space model and the state estimation problems}

Consider a time series $\{y_n\}$ described by the linear state-space model
\begin{align}
x_n &= F_n x_{n-1} + G_n v_n, \label{ssm-1} \\
y_n &= H_n x_n + w_n, \label{ssm-2}
\end{align}
where $x_n$ is a $k$-dimensional state vector.
The system noise $v_n$ and the observation noise $w_n$ are assumed to be
$\ell$-dimensional and one-dimensional white-noise sequences with density
functions $q_n(v)$ and $r_n(w)$, respectively.
The initial state vector $x_0$ is assumed to follow the density $p(x_0)$.

Let $Y_j$ denote the set of observations up to time $j$,
that is, $Y_j \equiv \{y_1,\ldots,y_j\}$.
The state estimation problem is to evaluate the conditional density
$p(x_n \mid Y_j)$ of the state $x_n$ given the observations $Y_j$ and the
initial density $p(x_0 \mid Y_0) \equiv p(x_0)$.
When $n>j$, $n=j$, and $n<j$, the problem is referred to as prediction,
filtering, and smoothing, respectively.

The linear state-space model \eqref{ssm-1}--\eqref{ssm-2} can be generalized
to a nonlinear non-Gaussian state-space model of the form
\begin{eqnarray}
x_n &=& F_n(x_{n-1}, v_n), \nonumber \\
y_n &=& H_n(x_n) + w_n, \label{ssm-3}
\end{eqnarray}
where $F_n(\cdot,\cdot)$ and $H_n(\cdot)$ are possibly nonlinear functions
of the state and the noise inputs.
A wide variety of problems in time-series analysis can be formulated and
analyzed using this nonlinear state-space framework
(\cite{KG 1996}, \cite{DFG 2001}).
This formulation can be further extended to a general state-space model
defined directly through conditional distributions.

\subsection{The Kalman filter and the smoother}

When all noise densities $q_n(v)$ and $r_n(w)$, as well as the initial state
density $p(x_0)$, are Gaussian, the conditional density
$p(x_n \mid Y_m)$ associated with the linear state-space model
\eqref{ssm-1}--\eqref{ssm-2} is also Gaussian.
In this case, the mean vector and covariance matrix can be computed
recursively using the Kalman filter and the fixed-interval smoothing
algorithm (\cite{AM 1979}).

Specifically, assume that
$q_n(v) \sim N(0,Q_n)$, $r_n(w) \sim N(0,R_n)$,
$p(x_0 \mid Y_0) \sim N(x_{0|0}, V_{0|0})$, and
$p(x_n \mid Y_m) \sim N(x_{n|m}, V_{n|m})$.
Then the Kalman filter is given as follows.

\noindent
\textbf{One-step-ahead prediction:}
\begin{eqnarray}
x_{n|n-1} &=& F_n x_{n-1|n-1}, \nonumber \\
V_{n|n-1} &=& F_n V_{n-1|n-1} F_n^{T} + G_n Q_n G_n^{T}.
\end{eqnarray}

\noindent
\textbf{Filtering:}
\begin{eqnarray}
K_n &=& V_{n|n-1} H_n^{T}
      \left(H_n V_{n|n-1} H_n^{T} + R_n \right)^{-1}, \nonumber \\
x_{n|n} &=& x_{n|n-1} + K_n \left(y_n - H_n x_{n|n-1} \right), \\
V_{n|n} &=& (I - K_n H_n) V_{n|n-1}. \nonumber
\end{eqnarray}

Using these filtering estimates, the smoothed density can be obtained by
the fixed-interval smoothing algorithm (\cite{AM 1979}).

\noindent
\textbf{Fixed-interval smoothing:}
\begin{eqnarray}
A_n &=& V_{n|n} F_n^{T} V_{n+1|n}^{-1}, \nonumber \\
x_{n|N} &=& x_{n|n} + A_n \left(x_{n+1|N} - x_{n+1|n} \right), \\
V_{n|N} &=& V_{n|n} +
A_n \left(V_{n+1|N} - V_{n+1|n} \right) A_n^{T}. \nonumber
\end{eqnarray}

\subsection{The non-Gaussian filter and the smoother}

For the nonlinear non-Gaussian state-space model \eqref{ssm-3}, the
conditional densities of the one-step-ahead predictor, the filter, and
the smoother satisfy the following recursive relations (\cite{Kitagawa 1987}).

\noindent
\textbf{One-step-ahead prediction:}
\begin{equation}
p(x_n \mid Y_{n-1})
= \int_{-\infty}^{\infty}
p(x_n \mid x_{n-1})\, p(x_{n-1} \mid Y_{n-1}) \, dx_{n-1}.
\end{equation}

\noindent
\textbf{Filtering:}
\begin{equation}
p(x_n \mid Y_n)
=
\frac{p(y_n \mid x_n)\, p(x_n \mid Y_{n-1})}
     {\int p(y_n \mid x_n)\, p(x_n \mid Y_{n-1}) \, dx_n}.
\end{equation}

\noindent
\textbf{Smoothing:}
\begin{equation}
p(x_n \mid Y_N)
=
p(x_n \mid Y_n)
\int_{-\infty}^{\infty}
\frac{p(x_{n+1} \mid Y_N)\, p(x_{n+1} \mid x_n)}
     {p(x_{n+1} \mid Y_n)} \, dx_{n+1}.
\end{equation}

\cite{Kitagawa 1987} and \cite{Kitagawa 1988} proposed a numerical algorithm for implementing the
non-Gaussian filter and smoother by approximating each density function
with a step function or a continuous piecewise linear function and
performing numerical integration.
This approach has been successfully applied to a wide range of problems,
including trend and volatility estimation, spectrum smoothing, smoothing
of discrete processes, and tracking problems
(\cite{KG 1996}; \cite{Kitagawa 2020}).

\section{Particle Filter and Smoothers}

The numerical-integration-based non-Gaussian filter and smoother
reviewed in the previous section are limited in practice to
low-dimensional state-space models, typically of third or fourth order.
To overcome this limitation, sequential Monte Carlo filtering and
smoothing methods, hereafter referred to as \emph{particle filters},
were developed.
In particle methods, each distribution appearing in the recursive
filtering and smoothing equations is approximated by a collection of
particles, which can be interpreted as random realizations from the
corresponding distribution (\cite{GSS 1993}; \cite{Kitagawa 1993}; \cite{Kitagawa 1996}).

\subsection{Forward Filter and Fixed-Lag Smoother}

Consider the state-space model
\begin{align}
x_n &\sim p(x_n \mid x_{n-1}), \\
y_n &\sim p(y_n \mid x_n),
\end{align}
where $y_n$ denotes the observation and $x_n$ the latent state.
The objective of marginal smoothing is to approximate, for each time
step $n$, the marginal posterior distribution
\begin{equation}
p(x_n \mid Y_N),
\end{equation}
conditioned on the entire observation sequence
$Y_N \equiv \{y_1,\ldots,y_N\}$.

\paragraph{Forward filtering.}
In the forward pass, a particle filter approximates the filtering
distribution as
\begin{equation}
p(x_n \mid Y_n)
\approx
\sum_{i=1}^m w_n^{(i)} \delta_{x_n^{(i)}}(x_n),
\end{equation}
where $\{x_n^{(i)}, w_n^{(i)}\}_{i=1}^m$ denote the particles and their
normalized weights.
These are obtained recursively through prediction and update steps.
The particle system at each time step is stored for use in subsequent
smoothing procedures.

\paragraph{ESS-based resampling in forward filtering.}
As time progresses, the importance weights
$\{w_n^{(i)}\}_{i=1}^m$ often become highly uneven, leading to weight
degeneracy, in which only a small number of particles carry most of the
probability mass.
A commonly used diagnostic for this phenomenon is the effective sample
size (ESS), defined by
\begin{equation}
\mathrm{ESS}_n
=
\frac{1}{\sum_{i=1}^m \left(w_n^{(i)}\right)^2},
\end{equation}
where the weights satisfy $\sum_{i=1}^m w_n^{(i)} = 1$.
The ESS takes values in $[1,m]$, with smaller values indicating more
severe degeneracy (\cite{LC 1998}; \cite{DFG 2001}).

To mitigate degeneracy while avoiding unnecessary resampling, an
ESS-based resampling strategy is employed.
Specifically, resampling is performed only when
\begin{equation}
\mathrm{ESS}_n < \alpha m,
\end{equation}
where $\alpha \in (0,1)$ is a user-specified threshold, typically chosen
between $0.3$ and $0.7$.
When $\mathrm{ESS}_n$ exceeds this threshold, resampling is skipped and
the weighted particle system is propagated to the next time step.

When resampling is triggered, a new set of particles
$\{x_n^{(i)}\}_{i=1}^m$ is generated by sampling with replacement from
the current particles according to their weights
$\{w_n^{(i)}\}_{i=1}^m$.
After resampling, all particles are assigned equal weights
$w_n^{(i)} = 1/m$.
This adaptive strategy balances variance reduction and particle
diversity and is therefore widely used in practical implementations of
particle filters.

\paragraph{Fixed-lag smoothing.}
It is interesting that particles approximating a smoothing
distribution can be obtained by a simple modification of the particle
filter (\cite{Kitagawa 1993}; \cite{Kitagawa 1996}).
Let $(s_{1|j}^{(i)},\ldots,s_{n|j}^{(i)})^T$ denote the $i$th realization
from the conditional joint distribution $p(x_1,\ldots,x_n \mid Y_j)$.
A fixed-lag smoothing algorithm is obtained by replacing Step~2(d) of
the filtering algorithm with the following procedure:

\vspace{2mm}
\hangindent=15mm
(d-S)\;
{\rm Generate
$\bigl\{(s_{n-L|n}^{(i)},\ldots,s_{n-1|n}^{(i)},s_{n|n}^{(i)})^T,
w_n^{(i)}\bigr\}$,
$i=1,\ldots,m$, from the previous state vectors
$\bigl\{(s_{n-L|n-1}^{(i)},\ldots,s_{n-1|n-1}^{(i)},p_n^{(i)})^T,
w_{n-1}^{(i)}\bigr\}$.
If $\mathrm{ESS}_n < \alpha m$, perform resampling of the state vectors.}

\vspace{2mm}

This procedure corresponds to an $L$-lag fixed-lag smoother rather than
a fixed-interval smoother (\cite{AM 1979}).
Increasing the lag $L$ generally improves the approximation of
$p(x_n \mid Y_{n+L})$ to $p(x_n \mid Y_N)$, but it may simultaneously
degrade the quality of the particle representation of
$p(x_n \mid Y_{n+L})$ due to path degeneracy (\cite{Kitagawa 1996}).
Since $p(x_n \mid Y_{n+L})$ often converges rapidly to $p(x_n \mid Y_N)$,
it is usually recommended to choose $L$ moderately.

\subsection{Backward marginal smoothing (FFBSm)}

The Forward Filter Backward Smoother in its marginal form (FFBSm)
computes the smoothing distributions $p(x_n \mid Y_N)$ directly, without
sampling complete state trajectories (\cite{DFG 2001}; \cite{GDW 2012}).
The method is based on the identity
\begin{equation}
p(x_n \mid Y_N)
=
p(x_n \mid Y_n)
\int
\frac{
p(x_{n+1} \mid x_n)\, p(x_{n+1} \mid Y_N)
}{
p(x_{n+1} \mid Y_n)
}
\,dx_{n+1}.
\end{equation}

Using the particle approximation obtained from the forward filter and
assuming
\begin{equation}
p(x_{n+1} \mid Y_N)
\approx
\sum_{j=1}^m \tilde w_{n+1}^{(j)}
\delta_{x_{n+1}^{(j)}}(x_{n+1}),
\end{equation}
the FFBSm recursion yields the following update for the smoothing
weights:
\begin{equation}
\tilde w_n^{(i)}
\propto
w_n^{(i)}
\sum_{j=1}^m
\tilde w_{n+1}^{(j)}
\frac{
p(x_{n+1}^{(j)} \mid x_n^{(i)})
}{
\sum_{k=1}^m
w_n^{(k)} p(x_{n+1}^{(j)} \mid x_n^{(k)})
},
\qquad i=1,\ldots,m.
\end{equation}
After normalization, the weighted particles
$\{x_n^{(i)},\tilde w_n^{(i)}\}_{i=1}^m$ provide an approximation of
$p(x_n \mid Y_N)$.

\paragraph{Computational complexity.}
The exact FFBSm recursion involves a double summation over particle
indices and has computational complexity $\mathcal{O}(m^2 N)$, which
can be prohibitive for large number of particles $m$.
Consequently, FFBSm is often used as a benchmark or reference smoother.

Various approximation techniques have been proposed to reduce the
computational cost, including local window truncation, kernel density
approximations, and importance-sampling-based methods.
These approaches reduce the complexity to
\begin{equation}
\mathcal{O}(mN)
\quad \text{or} \quad
\mathcal{O}(m m_s N),
\end{equation}
where $m_s \ll m$ denotes the size of a local subsample, at the cost of
introducing a controlled approximation error.

\paragraph{Discussion.}
Compared with backward-simulation-based smoothers (FFBSi), which
generate full state trajectories, FFBSm directly targets marginal
smoothing distributions and typically yields lower-variance estimates
of marginal expectations.
This property makes FFBSm particularly attractive for applications such
as parameter estimation via EM algorithms and offline state estimation,
where accurate marginal posterior distributions are required.

In summary, FFBSm provides a principled framework for marginal smoothing
in particle filters, offering exact but computationally demanding
formulations as well as efficient approximations that trade accuracy for
scalability.

\subsection{Subsamped FFBSm and Local Neighborhood Subsampling FFBSm}

To reduce the computational cost of the marginal Forward Filter Backward
Smoother (FFBSm), we consider a two approximation strategies based on
(i) truncating the state space using local neighborhoods and
(ii) performing subsampling within each local neighborhood.

\paragraph{Simple subsampling.}
As a first step toward reducing the computational cost of FFBSm, we
approximate the summation in equation~(17) by evaluating it over a subset
of indices.
Specifically, instead of computing the sum over the full index set
$J=\{1,\dots,m\}$, we restrict the summation to a subset
$\{j_1,\dots,j_{m_s}\}$ of size $m_s$.
For simplicity, we assume that $m_{\mathrm{int}} = m / m_s$ is an
integer.

Several straightforward strategies can be used to select $m_s$ indices
from $J$:
\begin{enumerate}
\item selecting $m_s$ equally spaced indices according to
      $j_i = j_0 + (i-1)m_{\mathrm{int}}$;
\item partitioning the interval $[1,m]$ into $m_s$ subintervals of length
      $m_{\mathrm{int}}$ and randomly sampling one index from each
      subinterval;
\item performing stratified sampling over $J$ using the weights
      $w_n^{(k)}$.
\end{enumerate}

In what follows, for the sake of simplicity, we focus on the first
strategy based on equally spaced sampling.
This approach is referred to as the \emph{subsampled FFBSm}
\textbf{(S--FFBSm)}.


\paragraph{Neighborhood definition.}
Let $V_n = x_{n+1} - x_n$ denote the state transition noise with cumulative
distribution function $F_V$.
For a prescribed tolerance level $\varepsilon > 0$, we define a distance
threshold $L_\varepsilon$ such that
\begin{equation}
\mathbb{P}(|V_n| > L_\varepsilon) = \varepsilon .
\end{equation}
Given a particle $x_{n+1}^{(j)}$ at time $n+1$, the corresponding local
neighborhood at time $n$ is defined as
\begin{equation}
\mathcal N_n(j)
=
\left\{
k \in \{1,\dots,m\} :
\|x_{n+1}^{(j)} - x_n^{(k)}\| \le L_\varepsilon
\right\}.
\end{equation}
In this study, we set $\varepsilon = 1/m$, so that the number of
particles $m$ controls the target size of the neighborhood.

For light-tailed transition models, such as the Gaussian distribution,
$L_\varepsilon$ grows only logarithmically with $m$, resulting in highly
localized neighborhoods.
In contrast, for heavy-tailed transition models, such as the Cauchy
distribution, $L_\varepsilon$ increases approximately linearly with $m$,
and the resulting neighborhood may cover a substantial portion of the
state space.
This difference reflects the fundamentally distinct tail behaviors of
the two noise models.

\paragraph{Subsampling within the neighborhood.}
Even when the neighborhood $\mathcal N_n(j)$ is relatively large,
additional computational savings can be achieved by approximating the
summations in the FFBSm recursion using a small subset of indices.
Specifically, for each $j$, we draw a subset
\begin{equation}
S_n(j) \subset \mathcal N_n(j), \qquad |S_n(j)| = m_{\text{small}},
\end{equation}
either uniformly or according to a prescribed sampling distribution.

Using this subset, the denominator in the FFBSm update is approximated
by
\begin{equation}
\hat D_j
=
\sum_{k \in S_n(j)}
\frac{w_n^{(k)}\,p(x_{n+1}^{(j)} \mid x_n^{(k)})}{\pi_{jk}},
\end{equation}
where $\pi_{jk}$ denotes the inclusion probability of index $k$ in
$S_n(j)$.
When the indices are sampled uniformly from $\mathcal N_n(j)$, we have
$\pi_{jk} = m_{\text{small}} / |\mathcal N_n(j)|$.
This Horvitz--Thompson type estimator provides an unbiased estimate of
the truncated denominator.

The smoothing weights are then updated according to
\begin{equation}
\tilde w_n^{(i)}
\propto
w_n^{(i)}
\sum_{j=1}^m
\tilde w_{n+1}^{(j)}
\frac{
p(x_{n+1}^{(j)} \mid x_n^{(i)})\,
\mathbf{1}\{i \in \mathcal N_n(j)\}
}{
\hat D_j
},
\qquad i=1,\dots,m.
\end{equation}

The same subsampling strategy can be applied to the numerator of the
FFBSm recursion, yielding an approximate update of the smoothing
weights.
The resulting algorithm has computational complexity
$\mathcal{O}(m\,m_{\text{small}}\,N)$, which is independent of the
possibly large size of the neighborhood $\mathcal N_n(j)$.
This approach is referred to as the \emph{neighborhood subsampled FFBSm}
\textbf{(NS--FFBSm)}.

\paragraph{Remark.}
For Gaussian transition models, the neighborhood $\mathcal N_n(j)$ is
typically small, and the additional subsampling introduces only minor
approximation error.
For Cauchy transitions, the neighborhood may span most of the particle
set; however, the subsampling step ensures that the computational cost
remains well controlled.
In this sense, the proposed approach provides a unified framework that
adapts naturally to both light-tailed and heavy-tailed transition models
while maintaining scalable computational complexity.

\section{Example: Smoothing Accuracy of a Trend Model}
\subsection{Models and Evaluation Criterion}

\begin{figure}[tbp]
\begin{center}
\includegraphics[width=80mm,angle=0,clip]{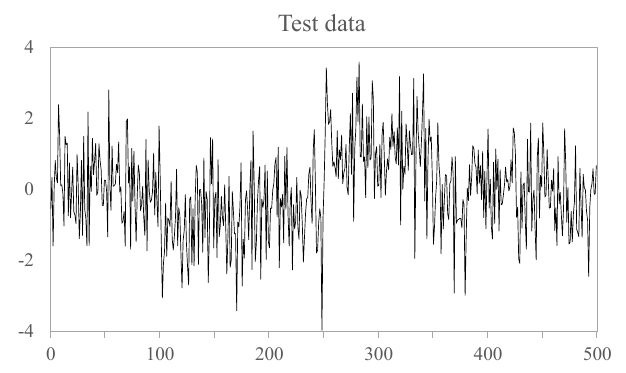}
\caption{Test data used for the empirical study (\cite{Kitagawa 1987}, \cite{Kitagawa 1996}, \cite{Kitagawa 2014}).}
\label{figure:Test_data}
\end{center}
\end{figure}

To evaluate the accuracy of particle smoothing distributions, we
consider the following first-order trend model:
\begin{eqnarray}
x_n &=& x_{n-1} + v_n, \nonumber \\
y_n &=& x_n + w_n,
\end{eqnarray}
where $y_n$ is the observed time series, $x_n$ is the trend component,
$v_n$ is the system noise, and $w_n$ is the observation noise.
We assume that the observation noise follows the Gaussian distribution
$w_n \sim N(0,\sigma^2)$, and that the system noise follows either the
Gaussian distribution $v_n \sim N(0,\tau^2)$ or the Cauchy distribution
$v_n \sim C(0,\tau^2)$.
Figure~\ref{figure:Test_data} shows the data set of length $N=500$ used
in the Monte Carlo experiments in this paper; the same data set was
also used in \cite{Kitagawa 1987}, \cite{Kitagawa 1996} and \cite{Kitagawa 2014}.

Following \cite{Kitagawa 2014}, we evaluate the smoothed distributions using
\begin{equation}
\mathrm{Dist}(D,\hat{D})
=
\sum_{n=1}^{500}\sum_{i=1}^{6400}
\left\{D(x_i,n)-\hat{D}(x_i,n)\right\}^2 \Delta x,
\end{equation}
where $D(x_i,n)$ and $\hat{D}(x_i,n)$ denote the ``true'' and estimated
smoothed probability density functions at time $n$, evaluated on grid
points $x_i$, $i=1,\ldots,6400$.
The grid is defined by $x_i = -8 + (i-1)\Delta x$ with $\Delta x=16/6400$.

\begin{figure}[tbp]
\begin{center}
\includegraphics[width=160mm,angle=0,clip]{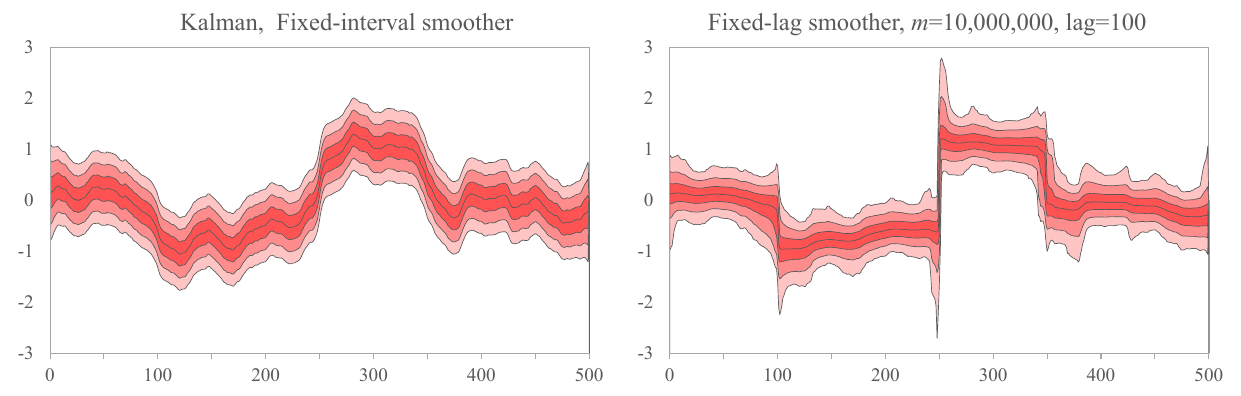}
\caption{``True'' smoothed distributions for the Gaussian and Cauchy models, obtained by the Kalman smoother and by a fixed-lag particle smoother with $m=10{,}000{,}000$ particles, respectively.}
\label{figure:True_smoother}
\end{center}
\end{figure}

Figure~\ref{figure:True_smoother} displays the mean and the $\pm1$ to
$\pm3$ standard-deviation bands of the ``true'' smoothed distributions
for the Gaussian and Cauchy models.
For the Gaussian model, the exact smoothing distribution is computed
using the Kalman filter and fixed-interval smoother.
For the Cauchy model, the ``true'' distribution is approximated by a
fixed-lag particle smoother with $10^7$ particles (\cite{Kitagawa 2014}).

\subsection{Fixed-lag Smoothing and FFBS}

\begin{table}[tbp]
\caption{Lag orders that minimize $\mathrm{Dist}(D,\hat{D})$ for the fixed-lag smoother, relative to the ``true'' fixed-interval smoother.}
\label{Tab_Accuracy_vs_LAG}
\begin{center}
\begin{tabular}{c|cccccc}
$m$  & $10^2$ & $10^3$ & $10^4$ & $10^5$ & $10^6$ & $10^7$ \\
\hline
Gaussian  & 16  & 22  & 27  & 32  & 43  & 53 \\
Cauchy    & 17  & 28  & 48  & 80  & 93  & 108
\end{tabular}
\end{center}
\end{table}

Table~\ref{Tab_Accuracy_vs_LAG} reports the lag length that minimizes
$\mathrm{Dist}(D,\hat{D})$ for particle numbers $m=10^k$, $k=2,\ldots,7$
(\cite{Kitagawa 2014}).
For both models, the optimal lag increases with $m$.
For the Gaussian model, the increase is moderate and a lag of around 50
is sufficient.
For the Cauchy model, the optimal lag grows more rapidly with $m$ and
exceeds 100 when $m=10^7$.

\begin{figure}[tbp]
\begin{center}
\includegraphics[width=160mm,angle=0,clip]{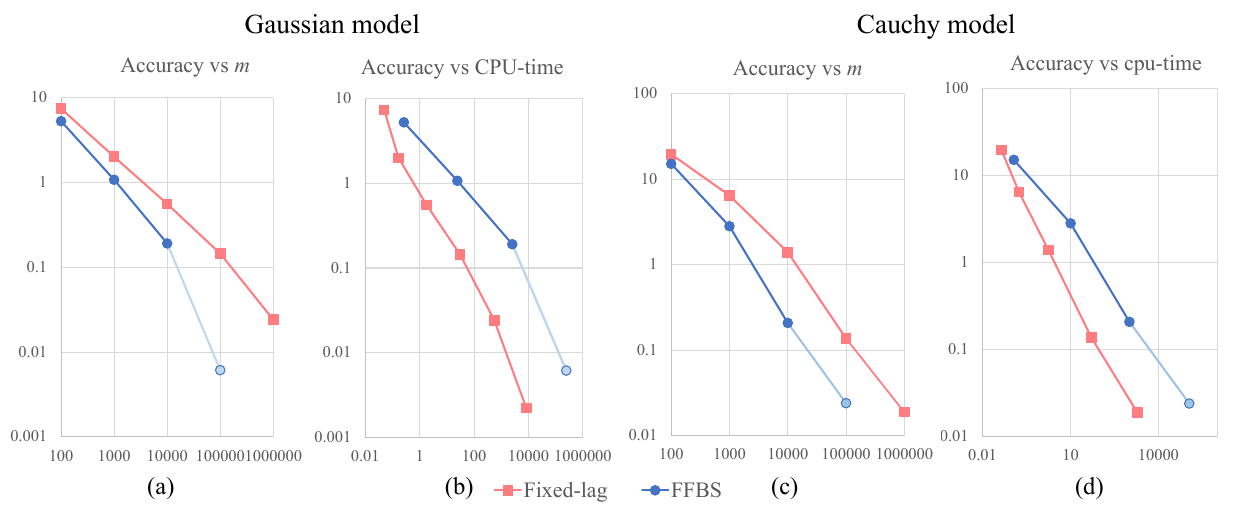}
\caption{Smoothing accuracy comparison for the Gaussian model.
(Left) Accuracy versus particle number $m$ (FFBS and fixed-lag smoothing).
(Right) Accuracy versus CPU time for the same experiments. FFBS with $m=10^5$
is computed only once and hence denoted by light blue.}
\label{figure:Fixed-lag_vs_FFBS}
\end{center}
\end{figure}

Figure~\ref{figure:Fixed-lag_vs_FFBS} and Table~\ref{Tab_Accuracy_Fixed-lag_vs_FFBS}
summarize the estimation accuracy and computational time of fixed-lag
smoothing and the Forward Filter Backward Smoother (FFBSm) for both the
Gaussian and Cauchy models, as the number of particles is varied from
$10^2$ to $10^7$.
Accuracy is evaluated by averaging $\mathrm{Dist}(D,\hat{D})$ over 100
independent runs with different random seeds.
For fixed-lag smoothing, the lag length is set to the optimal values in
Table~\ref{Tab_Accuracy_vs_LAG}.
All computations were performed on a PC equipped with an Intel Core
i7-8700 CPU running at 3.20--4.30\,GHz.

\begin{table}[tbp]
\begin{small}
\caption{Accuracy of the smoothing distributions and CPU time for the fixed-lag smoother and FFBSm.
The number of Monte Carlo repetitions is $\mathrm{NSIM}=100$.
(FFBSm with $m=100{,}000$ was performed only once.)
Values in parentheses are standard deviations.}
\label{Tab_Accuracy_Fixed-lag_vs_FFBS}
\begin{center}
\setlength{\tabcolsep}{1.5mm}
\begin{tabular}{c|cc|cc||cc|cc}
     & \multicolumn{4}{|c||}{Gaussian model} & \multicolumn{4}{c}{Cauchy model}\\
\cline{2-9}
     & \multicolumn{2}{|c|}{Accuracy} & \multicolumn{2}{c||}{CPU time (s)}
     & \multicolumn{2}{|c|}{Accuracy} & \multicolumn{2}{c}{CPU time (s)} \\
\cline{2-9}
$m$  & Fixed-lag & FFBSm & \makecell{Fixed\\[-1mm]lag} & FFBSm
     & Fixed-lag & FFBSm & \makecell{Fixed\\[-1mm]lag} & FFBSm \\
\hline
$10^2$ & 7.407(0.187)  &  5.269(0.201) &   0.05 &    0.24 & 19.585(0.656) & 15.135(0.600) &  0.04 &  0.12 \\
$10^3$ & 2.008(0.069)  &  1.074(0.062) &   0.17 &   23.1  &  6.459(0.070) &  2.816(0.151) &  0.18 &  10.1 \\
$10^4$ & 0.558(0.027)  &  0.191(0.010) &   1.79 & 2535.   &  1.396(0.026) &  0.208(0.011) &  1.75 &1027.  \\
$10^5$ & 0.145(0.013)  &  0.006(------) &  30.4  &245222  &  0.137(0.003) &  0.024(------)&  51.7 &112809 \\
$10^6$ & 0.024(0.004)  &                 & 551.   &       &  0.019(0.001) &               &  1001.&       \\
$10^7$ & 0.002(0.012)  &                 &8538    &       &  0.004(0.001) &               &  18246&
\end{tabular}
\end{center}
\end{small}
\end{table}

Table~\ref{Tab_Accuracy_Fixed-lag_vs_FFBS} suggests the following.
First, the qualitative dependence of accuracy and computational time on
$m$ is similar for the Gaussian and Cauchy models.
Second, both fixed-lag smoothing and FFBSm exhibit rapid improvement in
accuracy as $m$ increases, and FFBSm consistently achieves higher
accuracy than fixed-lag smoothing for the same $m$.
Third, the running time of FFBSm increases rapidly with $m$, reflecting
its quadratic computational complexity $\mathcal{O}(m^2)$.
Finally, when the methods are compared under the same computational
time budget, fixed-lag smoothing often attains superior accuracy.

Panels (a) and (c) of Figure~\ref{figure:Fixed-lag_vs_FFBS} show the
relationship between accuracy and $m$ for the Gaussian and Cauchy
models, respectively.
In both cases, the curves are approximately linear on a log--log scale.
The FFBSm results (blue) lie below the fixed-lag results (red),
indicating that FFBSm is more accurate at the same particle number.

Panels (b) and (d) compare accuracy against CPU time.
Here, the relative ordering is reversed: under the same computational
time budget, fixed-lag smoothing consistently achieves higher accuracy
than FFBSm.

\subsection{Effect of $\mathcal{O}(m)$ Approximation in FFBSm}

As discussed above, the exact FFBSm algorithm is computationally
prohibitive for large $m$.
We therefore focus on FFBSm and representative $\mathcal{O}(m)$
approximation strategies.

Specifically, we consider the following three approximations for the
summation in the FFBSm recursion:
\begin{itemize}
\item[(i)] replacing the full summation over all indices by a sum over
only $m_s$ equally spaced indices;
\item[(ii)] partitioning the index set $\{1,\dots,m\}$ into $m_s$
subintervals and sampling one index within each subinterval (stratified
sampling);
\item[(iii)] identifying a local neighborhood first and then applying
stratified sampling within that neighborhood.
\end{itemize}

The local neighborhood is defined according to the system noise model,
which is either Gaussian or Cauchy.
For each distribution, we determine $k=k_m$ such that
\begin{equation}
\mathbb{P}(|X|>k_m) = \frac{1}{m}.
\end{equation}
When the system noise has variance (or dispersion) parameter $\tau^2$,
the neighborhood associated with particle $x_{n+1}^{(j)}$ is defined as
\begin{equation}
\left[\, x_{n+1}^{(j)}-k_m\tau,\ \ x_{n+1}^{(j)}+k_m\tau \,\right].
\end{equation}

\begin{table}[bp]
\caption{Values of $k_m$ for Gaussian, Cauchy, and truncated Cauchy distributions.}
\label{Tab_optimal_k}
\begin{center}
\begin{tabular}{c|ccc|c|cc}
          & \multicolumn{3}{|c|}{Gaussian} & Cauchy & \multicolumn{2}{c}{Truncated Cauchy}\\
\cline{2-7}
$m$  &  $k_m$ & $\hat{k}_m$ & $k_m \tau_G$ &  $k_m$ & $\tilde{k}_m$ & $\tilde{k}_m\tau_{TC}$\rule{0mm}{5mm}  \\
\hline
$10^2$ & 2.5758 & 2.6630 & 0.2845 & 6.3657$\times 10$   & 6.137$\times 10$   & 0.3619 \\
$10^3$ & 3.2905 & 3.3668 & 0.3635 & 6.3662$\times 10^2$ & 4.629$\times 10^2$ & 2.7302 \\
$10^4$ & 3.8906 & 3.9593 & 0.4297 & 6.3662$\times 10^3$ & 1.339$\times 10^3$ & 7.8972 \\
$10^5$ & 4.4172 & 4.4802 & 0.4879 & 6.3662$\times 10^4$ & 1.651$\times 10^3$ & 9.7406 \\
$10^6$ & 4.8916 & 4.9502 & 0.5403 & 6.3662$\times 10^5$ & 1.691$\times 10^3$ & 9.9734
\end{tabular}
\end{center}
\end{table}

Table~\ref{Tab_optimal_k} reports the values of $k_m$ and $k_m\tau$ for
$m=10^2,\dots,10^6$ for the Gaussian, Cauchy, and truncated Cauchy
distributions.
For the Gaussian distribution, we also report the approximation
\[
\hat{k}_m \sim \sqrt{2\ln(2m)-\ln(\ln(2m))-\ln(2\pi)}
\]
(see \cite{Voutier 2010}).
For the truncated Cauchy distribution, the values $\tilde{k}_m$ are
obtained by Monte Carlo computation.

For the Gaussian distribution, $k_m$ ranges approximately from 2.5 to
4.9.
In our setting, the standard deviation of the Gaussian system noise is
$\tau_G=\sqrt{0.0122}=0.1105$, so $k_m\tau_G$ takes the values reported
in Table~\ref{Tab_optimal_k}.
In contrast, for the Cauchy distribution, $k_m$ increases rapidly with
$m$, so that the neighborhood becomes too wide to substantially reduce
the computational domain.

For the truncated Cauchy distribution with truncation limits $\pm10$,
Table~\ref{Tab_optimal_k} reports $k_m$ and $k_m\tau_{TC}$.
The values are obtained by Monte Carlo approximation.
In this setting, $k_m$ is small when $m=100$, whereas for $m\ge 10^4$ it
approaches 10, effectively corresponding to an untruncated range.

\begin{figure}[h]
\begin{center}
\includegraphics[width=150mm,angle=0,clip]{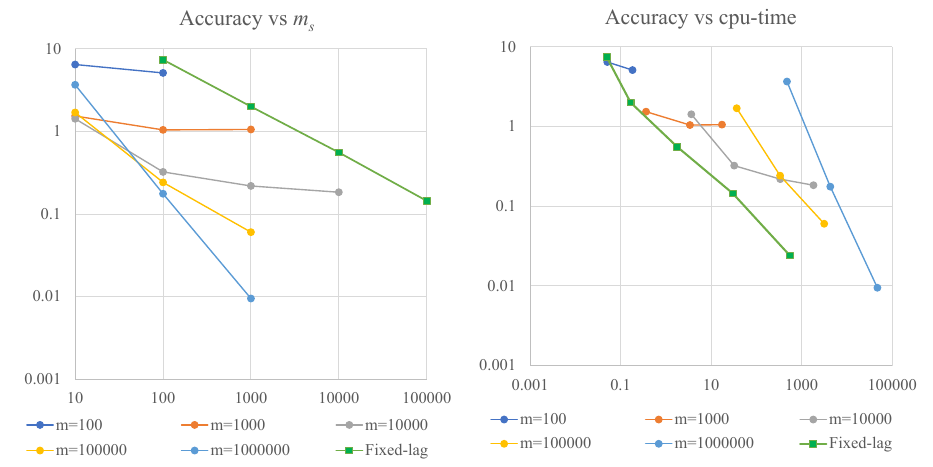}
\caption{Smoothing accuracy for the Gaussian model.
(Left) NS-FFBSm accuracy versus $m_s$ for different particle numbers $m$, 
with fixed-lag smoothing shown for comparison.
(Right) Accuracy versus CPU time for the same experiments.}
\label{figure:Accuracy_Gauss}
\end{center}
\end{figure}

Figure~\ref{figure:Accuracy_Gauss} illustrates the relationship between
smoothing accuracy and the choices of $m$ and $m_s$ for the Gaussian
model.
Numerical values are reported in Table~\ref{Tab_Accuracy_FFBS_vs_fast-FFBS, Gaussian model} in the Appendix.
In the left panel, the horizontal axis represents $m_s$ and the
vertical axis shows the accuracy measure $\mathrm{Dist}(D,\hat{D})$ for
NS-FFBSm.
The navy, red, gray, orange, and blue curves correspond to
$m=10^2$, $10^3$, $10^4$, $10^5$, and $10^6$, respectively.
For comparison, results for fixed-lag smoothing are also shown in green; in that
case, the horizontal axis is the particle number $m$.

The NS-FFBSm accuracy curves become nearly flat when $m_s$ is reduced to
approximately $m/100$, indicating that further improvements are
negligible beyond this point.
This suggests that choosing $m_s$ on the order of $m/100$ (or smaller)
is sufficient in practice for Gaussian transitions.
Fixed-lag smoothing is slightly less accurate than NS-FFBSm for each $m$,
which is likely attributable to degeneracy induced by repeated
resampling.

The right panel shows the same results plotted against CPU time.
Under a fixed computational budget, fixed-lag smoothing (green curve)
consistently achieves the best accuracy.
Although fixed-lag smoothing is less accurate than FFBSm and NS-FFBSm for the
same $m$, it requires only resampling of stored particles during
smoothing and therefore incurs relatively low additional cost.
As a result, fixed-lag smoothing can use a larger $m$ within the same
time budget and outperforms FFBSm and NS-FFBSm in terms of accuracy under equal
computational time.

\begin{figure}[tbp]
\begin{center}
\includegraphics[width=150mm,angle=0,clip]{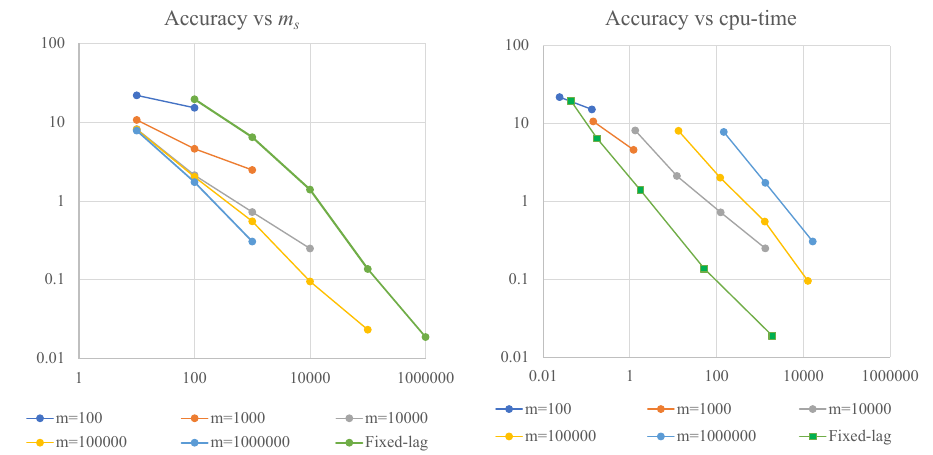}
\caption{Smoothing accuracy for the truncated Cauchy model.
(Left) Accuracy versus $m_s$ on a log--log scale.
(Right) Accuracy versus CPU time.}
\label{figure:Accuracy_Cauchy}
\end{center}
\end{figure}

Figure~\ref{figure:Accuracy_Cauchy} presents the corresponding results
for the truncated Cauchy model.
Numerical values are reported in Table~\ref{Tab_Accuracy_FFBS_vs_fast-FFBS, Cauchy} in the Appendix.
Unlike the Gaussian case, accuracy improves almost linearly with $m_s$
on a log--log scale.
This can be attributed to the fact that, except for very small $m$, the
neighborhood interval effectively covers most of the state space for
the Cauchy model.
Consequently, accurately approximating the smoothing distribution
requires larger $m_s$, and accuracy continues to improve as $m_s$
increases.
As in the Gaussian case, when $m_s=m$, FFBSm is more accurate than
fixed-lag smoothing for the same $m$, typically reducing the error by a
factor of several.

The right panel plots accuracy against CPU time.
Accuracy improves nearly linearly with computational time, and the
curves are approximately equally spaced.
Under the same computational budget, the choice $m_s=m$ achieves the
highest accuracy.
In this setting as well, fixed-lag smoothing can attain comparable or
higher accuracy under equal CPU time by employing a larger number of
particles.

Overall, Gaussian transitions allow efficient smoothing through strong
localization, so that FFBSm can achieve high accuracy with relatively
small $m_s$.
In contrast, heavy-tailed Cauchy transitions hinder localization and
require larger $m_s$ for substantial accuracy gains.
Fixed-lag smoothing, although less accurate than FFBSm-based methods at a
fixed particle number, often achieves superior accuracy under equal
computational time due to its substantially lower computational cost.

\subsection{Posterior Distributions Obtained by FFBSm}

In this subsection, we present illustrative examples of the posterior
distributions of the trend estimated by NS-FFBSm or S-FFBSm.

\begin{figure}[tbp]
\begin{center}
\includegraphics[width=160mm,angle=0,clip]{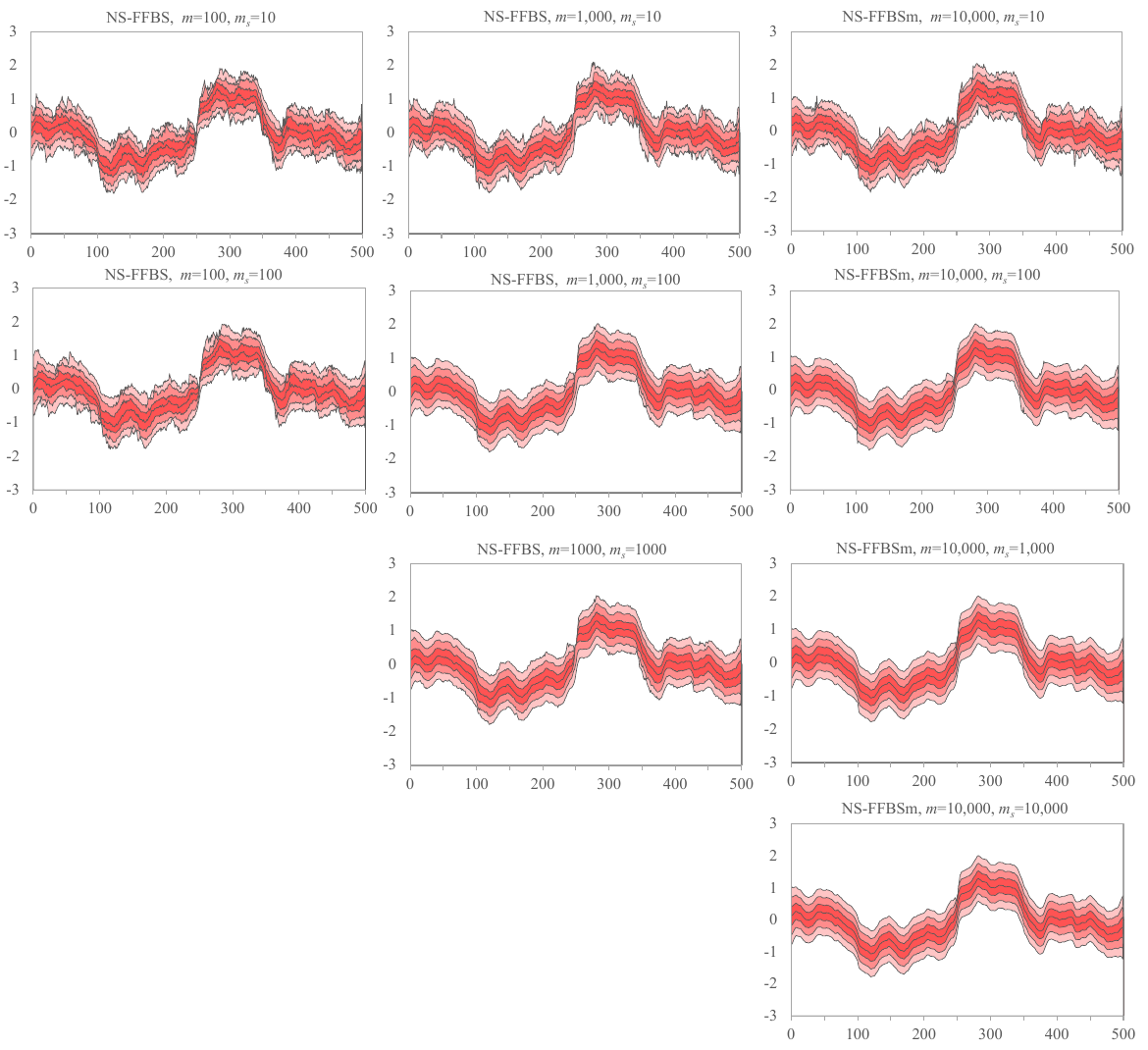}
\caption{Posterior distributions of the trend estimated by NS-FFBSm for the Gaussian model using local neighborhoods $\mathcal N_t(j)$.
Results are shown for $m=100,1000,$ and $10{,}000$ with $m_s=10,\dots,m$, including the posterior mean and the $\pm1$, $\pm2$, and $\pm3$ standard-deviation curves.}
\label{figure:Posterior_Gauss}
\end{center}
\end{figure}

Figure~\ref{figure:Posterior_Gauss} shows NS-FFBSm results for the Gaussian
model using the local neighborhood $\mathcal N_t(j)$, for
$m=100$, $1000$, and $10{,}000$, with $m_s=10,\dots,m$.
Even for the smallest configuration ($m=100$, $m_s=10$), both the
posterior mean and the $\pm1$ to $\pm3$ standard-deviation bands are
estimated reasonably well.
In particular, for $m\ge 1000$ and $m_s\ge 100$, the results are
comparable to those obtained with $m=m_s=10{,}000$ (bottom-right panel),
indicating no noticeable loss in estimation quality.

\begin{figure}[H]
\begin{center}
\includegraphics[width=160mm,angle=0,clip]{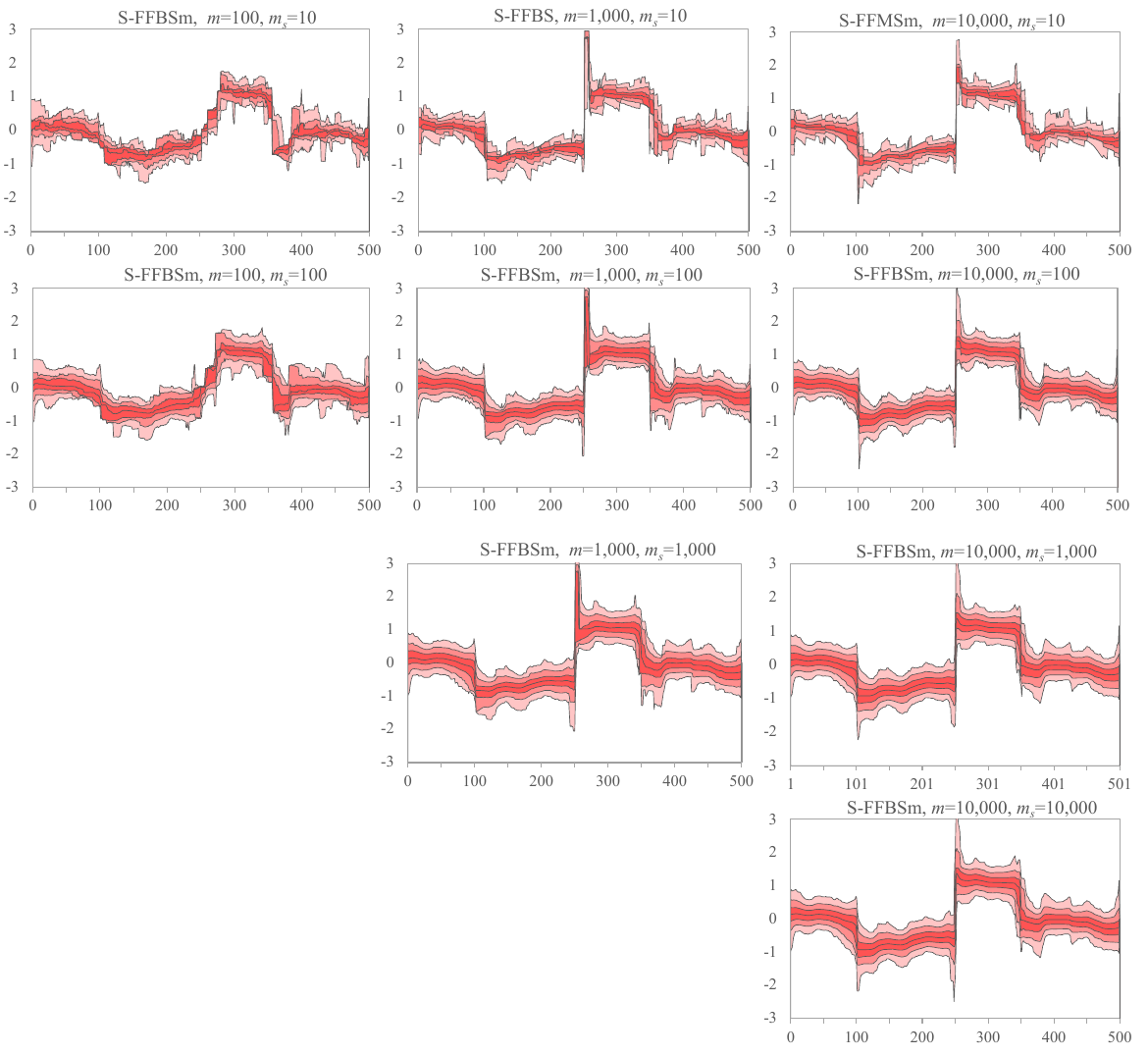}
\caption{Posterior distributions of the trend estimated by S-FFBSm for the Cauchy model.
Results are shown for different particle numbers $m$ and subsample sizes $m_s$, illustrating the effect of heavy-tailed system noise on the estimated posterior distributions.}
\label{figure:Posterior_Cauchy}
\end{center}
\end{figure}

Figure~\ref{figure:Posterior_Cauchy} shows S-FFBSm results for the Cauchy
model.
In contrast to the Gaussian case, trend jumps are not well captured
when $m=100$.
Moreover, even for larger $m$, noticeable irregularities and unnatural
discontinuities appear in the outer confidence bands when $m_s=10$.
Nevertheless, when $m\ge 1000$ and $m_s\ge 100$, the estimated posterior
distributions improve substantially.
In particular, $m=10{,}000$ with $m_s=100$ is sufficient to obtain a
reliable approximation in these experiments.

\section{Conclusion}

This paper investigated, through a trend-estimation example, the
practical accuracy--cost trade-offs of particle smoothing algorithms,
with particular emphasis on backward marginal smoothing (FFBSm) and its
computationally efficient approximations based on subsampling and local
neighborhoods.

The experiments lead to the following main conclusions.

\begin{enumerate}
\item \textbf{Accuracy versus particle number.}
For both Gaussian and heavy-tailed (truncated Cauchy) transition models,
the smoothing error measure $\mathrm{Dist}(D,\hat{D})$ decreases rapidly
as the number of particles $m$ increases. At a fixed particle number,
FFBSm/LS-FFBSm consistently achieves smaller error than fixed-lag smoothing,
reflecting the advantage of using information from the full data set in
a principled backward recursion.

\item \textbf{Cost dominates method ranking under equal CPU time.}
Despite its higher accuracy at fixed $m$, FFBSm (and exact FFBS) becomes
computationally prohibitive as $m$ increases due to its quadratic
complexity. When methods are compared under the same computational time
budget, fixed-lag smoothing often attains higher accuracy because it
allows substantially larger particle numbers within the same runtime.
This illustrates that the practical choice of smoother should be driven
by computational constraints rather than by asymptotic accuracy alone.

\item \textbf{Localization is decisive for efficient FFBSm.}
For Gaussian transitions, local neighborhoods are strongly localized,
and NS-FFBSm accuracy becomes nearly insensitive once the subsample size
$m_s$ reaches a small fraction of $m$ (on the order of $m/100$ in the
present study). Hence, substantial computational savings are possible
without noticeable loss of accuracy by combining FFBSm with
neighborhood restriction and subsampling.

\item \textbf{Heavy tails hinder localization and require larger $m_s$.}
For the (truncated) Cauchy transition model, the effective neighborhood
typically spans a large portion of the particle cloud, and the accuracy
continues to improve as $m_s$ increases. In this regime, aggressive
subsampling degrades the approximation, and the benefits of S-FFBSm
approximations are limited unless $m_s$ is chosen sufficiently large.

\item \textbf{Practical guidance.}
Taken together, these results suggest a simple rule of thumb: when the
state transition is light-tailed (e.g., Gaussian), FFBSm with local
neighborhoods and modest subsampling can provide an accurate and
computationally efficient smoother; when the transition is heavy-tailed,
fixed-lag smoothing with a larger particle number may be preferable
under a fixed computational budget.
\end{enumerate}

Overall, the single trend-estimation example clearly demonstrates that
the effectiveness of fast particle smoothing hinges on whether the
dynamics admit meaningful localization. Extending these findings to
higher-dimensional models and to other forms of state dynamics is an
important direction for future work.


\clearpage
\appendix
\section{Appendix}
\begin{table}[H]
\caption{Accuracy and cpu-time of Filter Distribution and Smoothing Distribution by Fixed-lag Smoothing and Backward Smoother. Gaussian distribution case. The values in parentheses indicate the standard deviation.}
\label{Tab_Accuracy_FFBS_vs_fast-FFBS, Gaussian model}
\begin{center}
\begin{footnotesize}
\begin{tabular}{c|c|c|ccc||cc|cc}
     &      & \multicolumn{4}{|c||}{Accuracy} & \multicolumn{4}{c}{Cpu time}\\
\cline{3-10}
   $m$  &   $m_s$ &Fixed-lag & FFBSm & S-FFBSm& NS-FFBSm&Fixed-lag& FFBSm & S-FFBSm& NS-FFBSm \\
\hline
 $10^2$ & $10  $ &(0.187)&(0.200)&  9.347  &  6.476 &        &        &   0.04 &    0.05 \\
        & $10^2$ & 7.407 & 5.269 &  5.350  &  5.127 &   0.05 &   0.24 &   0.25 &    0.18 \\
\hline 
 $10^3$ & $10  $ &       &       &  5.653  &  1.541 &        &        &   0.27 &    0.37 \\
        & $10^2$ &(0.069)&(0.062)&  1.556  &  1.047 &        &        &   2.49 &    3.42 \\
        & $10^3$ & 2.008 & 1.074 &  1.028  &  1.059 &   0.17 &  23.13 &  21.46 &   17.42 \\
\hline
 $10^4$ & $10  $ &       &       &  5.248  &  1.428 &        &        &   2.57 &    3.66 \\
        & $10^2$ &       &       &  0.947  &  0.324 &        &        &  24.84 &   32.47 \\
        & $10^3$ &(0.027)&(0.010)&  0.299  &  0.219 &        &        & 243.6  &  338.1 \\
        & $10^4$ & 0.558 & 0.191 &  0.2197 &  0.184 &   1.78 & 2535.  &2431.    & 18354. \\
\hline
 $10^5$ & $10  $ &       &       &  5.210  &  1.701 &        &        &  26.33 &   36.93 \\
        & $10^2$ &       &       &  0.849  &  0.242 &        &        & 257.83 &  335.2 \\
        & $10^3$ &       &       &  0.135  &  0.060 &        &        & 3235.  & 3160. \\
        & $10^4$ &(0.013)&(------)&  0.041 &  0.057 &        &        & 325453.&33571. \\   
        & $10^5$ & 0.144 & 0.006 &         &        &  30.48 & 245222 &        &\\
\hline
 $10^6$ & $10  $ &       &       &  3.679  &  3.877 &        &        & 356.9  &  476.1 \\
        & $10^2$ &       &       &  0.176  &  0.188 &        &        & 3257.  &  4314. \\
        & $10^3$ &(0.004)&       &  0.010  &  0.009 &        &        & 38117. &  47347.\\
        & $10^6$ & 0.024 &       &         &        & 551.06 &        &        &
\end{tabular}
\end{footnotesize}
\end{center}
%
%
%
\caption{Accuracy and cpu-time of Filter Distribution and Smoothing Distribution by Fixed-lag Smoothing and Backward Smoother. Truncated Cauchy distribution case. The values in parentheses indicate the standard deviation.}
\label{Tab_Accuracy_FFBS_vs_fast-FFBS, Cauchy}
\begin{center}
\begin{footnotesize}
\begin{tabular}{c|l|c|cc||c|cc}
     &      &\multicolumn{3}{|c||}{Accuracy} & \multicolumn{3}{c}{Cpu time}\\
\cline{3-8}
   $m$  &   $m_s$ &Fixed-lag & FFBSm & S-FFBSm& Fixed-lag& FFBSm & S-FFBSm \\
\hline
  $10^2$  &     10 &(0.656)& (0.600)& 21.916(0.804) &       &        &   0.02  \\
          & $10^2$ &19.585 & 15.135 & 15.2465(0.552)       &  0.04 &   0.12 &   0.13  \\
\hline 
          &     10 &       &        & 10.688(0.287) &       &        &   0.14  \\
  $10^3$  & $10^2$ &(0.230)&(0.151) &  4.603(0.218) &       &        &   1.22  \\
          & $10^3$ & 6.459 & 2.816  &  2.477(0.128) &  0.18 &  10.11 &  13.45  \\
\hline
          &     10 &       &        &  8.315(0.072) &       &        &   1.33  \\
  $10^4$  & $10^2$ &       &        &  2.128(0.059) &       &        &  12.13  \\
          & $10^3$ &(0.026)&(0.010) &  0.673(0.032) &       &        & 123.94  \\
          & $10^4$ & 1.396 & 0.208  &  0.250(0.014) &  1.75 &1027.70 &1325.23  \\
\hline
          &     10 &       &        &  8.000(0.045) &       &        &  13.21  \\
          & $10^2$ &       &        &  1.959(0.041) &       &        & 120.63  \\
  $10^5$  & $10^3$ &       &        &  0.552(0.102) &       &        &1291.27  \\
          & $10^4$ &(0.003)&(------)&  0.087(0.014) &       &        &12705.0  \\   
          & $10^5$ & 0.137 & 0.024  &  0.023(------)& 51.74 &101319  &130620   \\   
\hline
          &     10 &       &        &  7.822(0.084) &       &        & 146.17  \\
  $10^6$  & $10^2$ &       &        &  1.738(0.036) &       &        &1328.69  \\
          & $10^3$ &(0.001)&        &  0.306(0.059) &       &        &16409.00 \\
          & $10^6$ & 0.019 &        &          &1001.12&        &
\end{tabular}
\end{footnotesize}
\end{center}
\end{table}

\end{document}